\pdfoutput=1

\documentclass[11pt]{article}

\usepackage{EMNLP2022}

\usepackage{times}
\usepackage{latexsym}

\usepackage[T1]{fontenc}

\usepackage[utf8]{inputenc}

\usepackage{microtype}

\usepackage{inconsolata}

\usepackage{graphicx}
\usepackage{subfigure}
\usepackage{amsmath}
\usepackage{multirow}
\usepackage{makecell}
\usepackage{hyperref}
\usepackage{booktabs}
\usepackage{enumitem}
\let\sss= \scriptscriptstyle

\title{Tiny-NewsRec: Effective and Efficient PLM-based News Recommendation}

\author{
  Yang Yu\textsuperscript{\rm 1}, Fangzhao Wu\textsuperscript{\rm 2}, Chuhan Wu\textsuperscript{\rm 3}, Jingwei Yi\textsuperscript{\rm 1} and Qi Liu\textsuperscript{\rm 1} \\
  \textsuperscript{\rm 1}University of Science and Technology of China\\
  \textsuperscript{\rm 2}Microsoft Research Asia\ \textsuperscript{\rm 3}Tsinghua University\\
  \texttt{\{yflyl613, yjw1029\}@mail.ustc.edu.cn}\\
  \texttt{\{wufangzhao, wuchuhan15\}@gmail.com, qiliuql@ustc.edu.cn}
}

\begin{document}
\maketitle

\begin{abstract}
    News recommendation is a widely adopted technique to provide personalized news feeds for the user.
    Recently, pre-trained language models (PLMs) have demonstrated the great capability of natural language understanding and benefited news recommendation via improving news modeling.
    However, most existing works simply finetune the PLM with the news recommendation task, which may suffer from the known domain shift problem between the pre-training corpus and downstream news texts.
    Moreover, PLMs usually contain a large volume of parameters and have high computational overhead, which imposes a great burden on low-latency online services.
    In this paper, we propose Tiny-NewsRec, which can improve both the effectiveness and the efficiency of PLM-based news recommendation.
    We first design a self-supervised domain-specific post-training method to better adapt the general PLM to the news domain with a contrastive matching task between news titles and news bodies.
    We further propose a two-stage knowledge distillation method to improve the efficiency of the large PLM-based news recommendation model while maintaining its performance.
    Multiple teacher models originated from different time steps of our post-training procedure are used to transfer comprehensive knowledge to the student in both its post-training and finetuning stage.
    Extensive experiments on two real-world datasets validate the effectiveness and efficiency of our method.
\end{abstract}
\section{Introduction}
With the explosion of information, massive news is published on online news platforms such as Microsoft News and Google News~\cite{das2007google, lavie2010user}, which can easily get the users overwhelmed when they try to find the information they are interested in~\cite{okura2017embedding,2020zhi}.
Many personalized news recommendation methods have been proposed to alleviate the information overload problem for users~\cite{wang2018dkn,wu2019npa,zhu2019dan,hu2019graph}.
Since news articles usually contain abundant textual content, learning high-quality news representations from news texts is one of the most critical tasks for news recommendation~\cite{wu2020mind}.
As pre-trained language models (PLMs) have been proved to be powerful in text modeling and have empowered various NLP tasks~\cite{devlin2019bert, liu2019roberta}, a few recent works delve into employing PLMs for better news modeling in news recommendation~\cite{wu2021empower,rmbert2021,zhang2021unbert}.
For example, \citet{wu2021empower} propose to replace shallow NLP models such as CNN and attention network with the PLM to capture the deep contexts in news texts.
However, these methods simply finetune the PLM with the news recommendation task, which may be insufficient to cope with the domain shift problem between the generic pre-training corpus and downstream news texts~\cite{gururangan2020adapt,madan-etal-2021-tadpole}.
Moreover, PLMs usually have a large number of parameters.
For example, the BERT-base model~\cite{devlin2019bert} contains 12 layers with 110M parameters.
Deploying these PLM-based news recommendation models to provide low-latency online services requires extensive computational resources.

In this paper, we propose a Tiny-NewsRec approach to improve both the effectiveness and the efficiency of PLM-based news recommendation\footnote{The source code and data of our Tiny-NewsRec are available at \href{https://github.com/yflyl613/Tiny-NewsRec}{https://github.com/yflyl613/Tiny-NewsRec}.}.
In our approach, we first utilize the natural matching relation between different parts of a news article and design a self-supervised domain-specific post-training method to better adapt the general PLM to the news domain.
The PLM-based news encoder is trained with a contrastive matching task between news titles and news bodies to make it better capture the semantic information in news texts and generate more discriminative representations, which are beneficial to both news content understanding and user interest matching in the following news recommendation task.
In addition, we propose a two-stage knowledge distillation method to compress the large PLM-based model while maintaining its performance\footnote{We focus on task-specific knowledge distillation.}.
Domain-specific knowledge and task-specific knowledge are transferred from the teacher model to the student in its post-training stage and finetuning stage respectively.
Besides, multiple teacher models originated from different time steps of our post-training procedure are used to provide comprehensive guidance to the student model in both stages.
For each training sample, we adaptively weight these teacher models based on their performance, which allows the student model to always learn more from the best teacher.
Extensive experiment results on two real-world datasets show that our approach can reduce the model size by 50\%-70\% and accelerate the inference speed by 2-8 times while achieving better performance.
The main contributions of our paper are as follows:
\begin{itemize}[leftmargin=*,itemsep=2pt,parsep=0pt,topsep=4pt]
    \item We propose a Tiny-NewsRec approach to improve both the effectiveness and efficiency of PLM-based news recommendation.
    \item We propose a self-supervised domain-specific post-training method which trains the PLM with a contrastive matching task between news titles and news bodies before the task-specific finetuning to better adapt it to the news domain.
    \item We propose a two-stage knowledge distillation method with multiple teacher models to compress the large PLM-based model.
    \item Extensive experiments on two real-world datasets validate that our method can effectively improve the performance of PLM-based news recommendation models while reducing the model size by a large margin.
\end{itemize}
\section{Related Work}
\subsection{PLM-based News Recommendation}
With the great success of pre-trained language models (PLMs) in multiple NLP tasks, many researchers have proposed to incorporate the PLM in news recommendation and have achieved substantial gain \cite{zhang2021unbert,rmbert2021,wu2021empower}.
For example, \citet{zhang2021unbert} proposed UNBERT, which utilizes the PLM to capture multi-grained user-news matching signals at both word-level and news-level.
\citet{wu2021empower} proposed a state-of-the-art PLM-based news recommendation method named PLM-NR, which instantiates the news encoder with a PLM to capture the deep semantic information in news texts and generate high-quality news representations.
However, these methods simply finetune the PLM with the news recommendation task, the supervision from which may be insufficient to fill the domain gap between the generic pre-training corpus and downstream news texts~\cite{gururangan2020adapt,madan-etal-2021-tadpole}.
Besides, PLMs usually contain a large number of parameters and have high computational overhead.
Different from these methods, our approach can better mitigate the domain shift problem with an additional domain-specific post-training task and further reduce the computational cost with a two-stage knowledge distillation method.

\subsection{Domain Adaptation of the PLM}
Finetuning a PLM has become a standard procedure for many NLP tasks~\cite{devlin2019bert, 2020t5}.
These models are first pre-trained on large generic corpora (e.g., BookCorpus and Wikipedia) and then finetuned on the downstream task data.
Even though this paradigm has achieved great success, it suffers from the known domain shift problem between the pre-training and downstream corpus~\cite{howard-ruder-2018-universal, lee2019biobert, beltagy-etal-2019-scibert}.
A technique commonly used to mitigate this problem is continuing to pre-train the general PLM on additional corpora related to the downstream task~\cite{logeswaran-etal-2019-zero, chakrabarty-etal-2019-imho,han-eisenstein-2019-unsupervised}.
For example, \citet{gururangan2020adapt} proposed domain-adaptive pre-training (DAPT) and task-adaptive pre-training (TAPT), which further pre-trains the PLM on a large corpus of unlabeled domain-specific text and the training text set for a given task before the task-specific finetuning, respectively.
Instead of continued pre-training, we utilize the natural matching relation between different parts of a news article and design a domain-specific post-training method with a contrastive matching task between news titles and news bodies.
It can make the PLM better capture the high-level semantic information in news texts and generate more discriminative news representations, which are beneficial for news recommendation.

\subsection{PLM Knowledge Distillation}
Knowledge distillation (KD) is a technique that aims to compress a heavy teacher model into a lightweight student model while maintaining its performance~\cite{hinton2015distill}.
In recent years, many works explore compressing large-scale PLMs via KD~\cite{sun2019bertpkd,wang2020minilm,sun2020mobilebert,xu2020theseus}.
For example,~\citet{jiao2020tinybert} proposed TinyBERT, which lets the student model imitate the intermediate and final outputs of the teacher model in both the pre-training and finetuning stages.
There are also a few works that aim to distill the PLM for specific downstream tasks~\cite{lu2021twinbert,wu2021newsbert}.
For example,~\citet{wu2021newsbert} proposed NewsBERT for intelligent news applications.
A teacher-student joint distillation framework is proposed to collaboratively learn both teacher and student models.
Considering that the guidance provided by a single teacher may be limited or even biased, some works propose to conduct KD with multiple teacher models~\cite{adaptive2020,wu2021teacher}.
However, all these works neglect the potential domain gap between the pre-training corpus and the downstream task domain.
To our best knowledge, we are the first to conduct KD during the domain adaptation of PLMs.
Both domain-specific and task-specific knowledge are transferred to the student model in our two-stage knowledge distillation method.
Besides, multiple teacher models are used to provide more comprehensive guidance to the student in both stages.
\section{Methodology}
In this section, we introduce the details of our Tiny-NewsRec method.
We first briefly introduce the structure of our PLM-based news recommendation model.
Then we introduce the design of our self-supervised domain-specific post-training method and the framework of our two-stage knowledge distillation method.
Some notations used in the paper are listed in Table~\ref{notation}.

\subsection{News Recommendation Model}

We first introduce the structure of the PLM-based news recommendation model used in our Tiny-NewsRec.
As shown in Fig.~\hyperref[fig.framework]{1(b)}, it consists of three major components, i.e., a news encoder, a user encoder, and a click prediction module.
The news encoder aims to learn the news representation from news texts.
Following the state-of-the-art PLM-based news recommendation method~\cite{wu2021empower}, we use a PLM to capture the deep context in news texts and an attention network to aggregate the output of the PLM.
The user encoder aims to learn the user representation from the representations of the user's last $L$ clicked news, i.e., $[\boldsymbol{n}_1, \boldsymbol{n}_2, ..., \boldsymbol{n}_{\sss L}]$.
Following~\citet{wu2019naml}, we implement it with an attention network to select important news from the user's historical interactions.
In the click prediction module, we take the dot product of the candidate news representation $\boldsymbol{n}_c$ and the target user representation $\boldsymbol{u}$ as the predicted score $\hat{y}_{\sss \rm{FT}}$.
It is noted that our Tiny-NewsRec is decoupled from the structure of the news recommendation model.
Other PLM-based news recommendation models~\cite{rmbert2021,amm2021,zhang2021unbert} can also be adopted.

\subsection{Domain-specific Post-training}
\begin{table}[!t]
  \centering
  \setlength{\belowcaptionskip}{-5pt}
  \resizebox{\linewidth}{!}{
    \begin{tabular}{ll}
      \specialrule{1.5pt}{0pt}{1.5pt}
      \multicolumn{1}{l}{Notation}                 & \multicolumn{1}{l}{Explanation}                   \\ \specialrule{0.5pt}{0pt}{1.5pt}
      $\boldsymbol{h}_{nb}$                        & News body representation                          \\
      $\boldsymbol{h}_{nt}$                        & News title representation                         \\
      $\boldsymbol{n}$                             & News representation                               \\
      $\boldsymbol{u}$                             & User representation                               \\
      $\operatorname{CE}(\cdot,\cdot)$             & Cross-Entropy loss function                       \\
      $\operatorname{MSE}(\cdot,\cdot)$            & Mean-Squared Error loss function                  \\
      ${}^{\sss (t_{\scalebox{0.5}{\textit{i}}})}$ & Outputs or parameters of the $i$-th teacher model \\
      ${}^{\sss (s)}$                              & Outputs or parameters of the student model        \\
      $\rm{FT}$                                    & Abbreviation for "Finetune"                       \\
      $\rm{DP}$                                    & Abbreviation for "Domain-specific Post-train"     \\
      \specialrule{1.5pt}{1pt}{0pt}
    \end{tabular}
  }
  \caption{Some notations used in this paper.}\label{notation}
\end{table}

\begin{figure*}[!t]
  \centering
  \includegraphics[width=\linewidth]{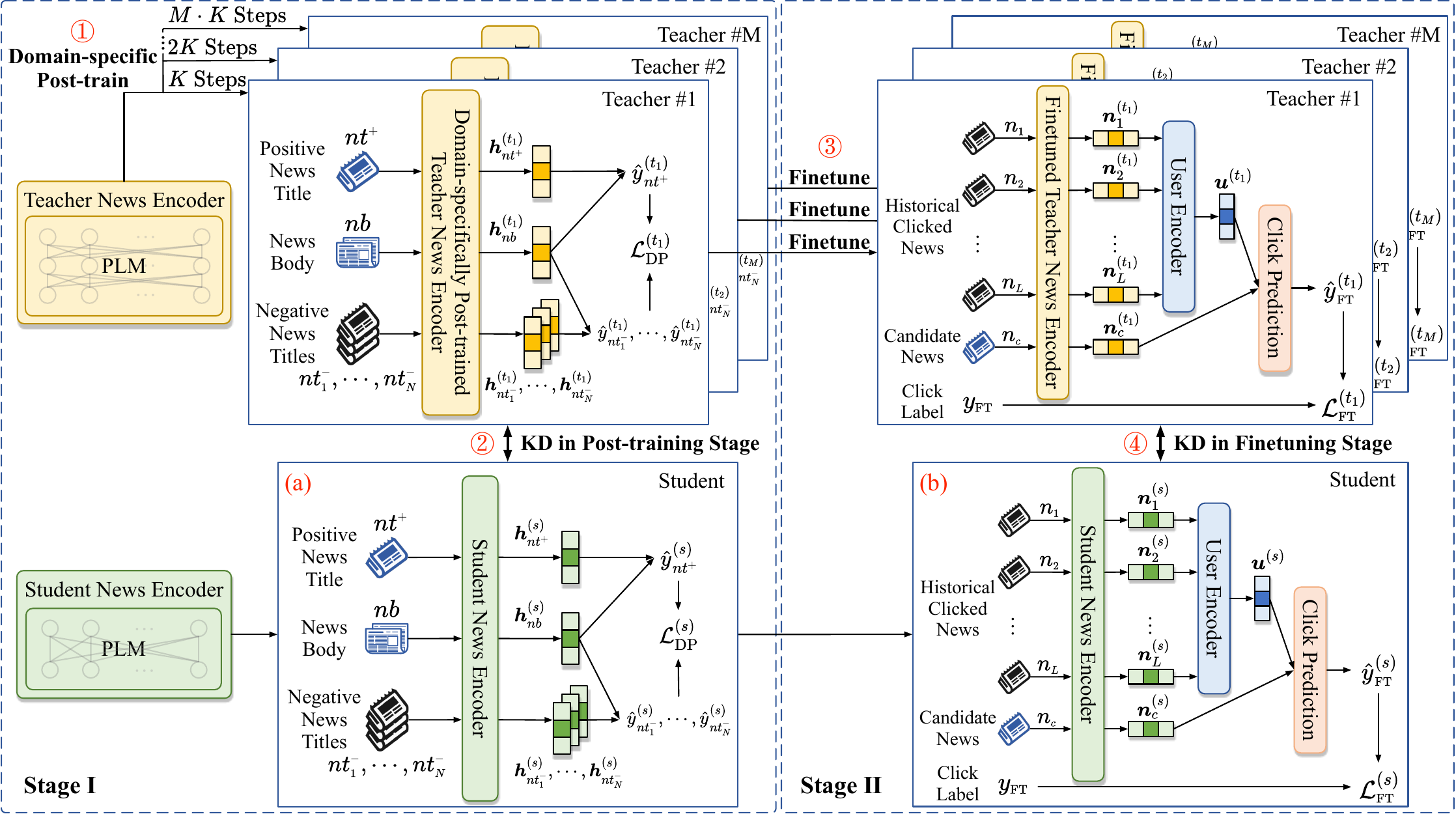}
  \caption{The framework of Tiny-NewsRec}
  \label{fig.framework}
\end{figure*}

Since directly finetuning the PLM with the downstream news recommendation task may be insufficient to fill the domain gap between the general corpus and news texts~\cite{gururangan2020adapt,madan-etal-2021-tadpole}, we propose to conduct domain-specific post-training to the PLM before the task-specific finetuning.
Considering the natural matching relation between different parts of a news article, we design a self-supervised contrastive matching task between news titles and news bodies.
The model framework for this task is shown in Fig.~\hyperref[fig.framework]{1(a)}.

Given a news article, we regard its news body $nb$ as the anchor and take its news title $nt^+$ as the positive sample.
We randomly select $N$ other news titles $[nt_1^-, nt_2^-,\cdots, nt_{\sss N}^-]$ from the news pool as negative samples.
We use the PLM-based news encoder to get the news body representation $\boldsymbol{h}_{nb}$ and these news title representations $[\boldsymbol{h}_{nt^{\scalebox{0.5}{+}}}, \boldsymbol{h}_{nt_1^{\scalebox{1}{-}}}, \boldsymbol{h}_{nt_2^{\scalebox{1}{-}}},\cdots, \boldsymbol{h}_{nt_{\scalebox{0.5}{\textit{N}}}^{\scalebox{1}{-}}}]$.
We adopt the InfoNCE loss~\cite{oord2018cpc} as the contrastive loss function. It is formulated as follows:
\begin{equation*}
  \mathcal{L}_{\sss \rm{DP}}= -\operatorname{log}\frac{\exp({\hat{y}_{nt^{\scalebox{0.5}{+}}}})}{\exp({\hat{y}_{nt^{\scalebox{0.5}{+}}}}) + \sum_{i=1}^{\sss N}\exp({\hat{y}_{nt^{\scalebox{1}{-}}_{\scalebox{0.5}{\textit{i}}}}})},
\end{equation*}
where $\hat{y}_{nt^{\scalebox{0.5}{+}}}=\boldsymbol{h}_{nb}^{\sss \rm{T}}\boldsymbol{h}_{nt^{\scalebox{0.5}{+}}}$ and $\hat{y}_{nt^{\scalebox{1}{-}}_{\scalebox{0.5}{\textit{i}}}}=\boldsymbol{h}_{nb}^{\sss \rm{T}}\boldsymbol{h}_{nt^{\scalebox{1}{-}}_{\scalebox{0.5}{\textit{i}}}}$.
As proved by~\citet{oord2018cpc}, minimizing $\mathcal{L}_{\sss \rm{DP}}$ can maximize the lower bound of the mutual information between $\boldsymbol{h}_{nb}$ and $\boldsymbol{h}_{nt^{\scalebox{0.5}{+}}}$.
Therefore, the post-trained PLM-based news encoder can better capture and match the high-level semantic information in news texts.
It will generate more similar representations for related texts (i.e., the news body and its corresponding news title) and distinguish them from the others, which can also ease the anisotropy problem of the sentence representation generated by the PLM~\cite{jun2019degeneration,ethayarajh2019contextual,li2020sentence}.
Thus, our proposed domain-specific post-training method is beneficial to both news understanding and user interest matching in the following news recommendation task.

\subsection{Two-stage Knowledge Distillation}
To achieve our goal of efficiency, we further propose a two-stage knowledge distillation method, whose framework is shown in Fig.~\ref{fig.framework}.
In our framework, the lightweight student model is trained to imitate the large teacher model in both its post-training stage and finetuning stage.
Besides, multiple teacher models originated from different time steps of our post-training procedure are used to transfer more comprehensive knowledge to the student model in both stages.

In Stage I, we first conduct domain-specific post-training towards the teacher PLM-based news encoder (Step 1).
During the post-training procedure, a copy of the current teacher news encoder is saved every $K$ steps after convergency and we save $M$ teacher models in total.
Then we use these teacher models to transfer comprehensive domain-specific knowledge to the student model during its post-training (Step 2).
Since these teacher models at different time steps may have different performance on an input sample, we assign an adaptive weight to each teacher for each training sample, which is measured by the cross-entropy loss between its predicted scores $\hat{\boldsymbol{y}}_{\sss \rm{DP}}^{\sss (t_{\scalebox{0.5}{\textit{i}}})}=[\hat{y}^{\sss (t_{\scalebox{0.5}{\textit{i}}})}_{nt^{\scalebox{0.5}{+}}}, \hat{y}^{\sss (t_{\scalebox{0.5}{\textit{i}}})}_{nt_{\scalebox{0.4}{1}}^{\scalebox{1}{-}}},\hat{y}^{\sss (t_{\scalebox{0.5}{\textit{i}}})}_{nt_{\scalebox{0.4}{2}}^{\scalebox{1}{-}}},\cdots,\hat{y}^{\sss (t_{\scalebox{0.5}{\textit{i}}})}_{nt_{\scalebox{0.4}{\textit{N}}}^{\scalebox{1}{-}}}]$ and the ground-truth label $y_{\sss \rm{DP}}$.
Denote the weight of the $i$-th teacher model on a given sample as $\alpha^{\sss (t_{\scalebox{0.5}{\textit{i}}})}$, it is formulated as follows:
\begin{equation*}
  \alpha^{\sss (t_{\scalebox{0.5}{\textit{i}}})} = \frac{\exp(-\operatorname{CE}(\hat{\boldsymbol{y}}_{\sss \rm{DP}}^{\sss (t_{\scalebox{0.5}{\textit{i}}})}, y_{\sss \rm{DP}}))}{\sum_{j=1}^{\sss M}{\exp(-\operatorname{CE}(\hat{\boldsymbol{y}}_{\sss \rm{DP}}^{\sss (t_j)}, y_{\sss \rm{DP}}))}}.
\end{equation*}
To encourage the student model to make similar predictions to the best teacher model, we use a distillation loss to regularize its output soft labels, which is formulated as follows:
\begin{equation*}
  \mathcal{L}_{\sss \rm{DP}}^{\sss \rm{distill}} = T_{\sss \rm{DP}}^{\sss 2}\cdot\operatorname{CE}(\sum_{i=1}^{M}{\alpha^{\sss (t_{\scalebox{0.5}{\textit{i}}})}\hat{\boldsymbol{y}}_{\sss {\rm{DP}}}^{\sss (t_{\scalebox{0.5}{\textit{i}}})}}/T_{\sss \rm{DP}}, \hat{\boldsymbol{y}}_{\sss \rm{DP}}^{\sss (s)}/T_{\sss \rm{DP}}).
\end{equation*}
$T_{\sss \rm{DP}}$ is a temperature hyper-parameter that controls the smoothness of the predicted probability distribution of the teacher models.
Besides, since we expect the representations generated by the student model and these teacher models to be similar in a unified space, we propose to apply an additional embedding loss to align these representations.
The embedding loss between the $i$-th teacher model and the student model is formulated as follows:
\begin{align*}
  \mathcal{L}^{\sss {\rm{emb}_{\scalebox{0.5}{\textit{i}}}}}_{\sss \rm{DP}}= & \operatorname{MSE}(\boldsymbol{W}^{\sss (t_{\scalebox{0.5}{\textit{i}}})}\boldsymbol{h}_{nt}^{\sss (t_{\scalebox{0.5}{\textit{i}}})}+\boldsymbol{b}^{\sss (t_{\scalebox{0.5}{\textit{i}}})},\boldsymbol{h}_{nt}^{\sss (s)})+  \\
                                                                             & \operatorname{MSE}(\boldsymbol{W}^{\sss (t_{\scalebox{0.5}{\textit{i}}})}\boldsymbol{h}_{nb}^{\sss (t_{\scalebox{0.5}{\textit{i}}})}+\boldsymbol{b}^{\sss (t_{\scalebox{0.5}{\textit{i}}})}, \boldsymbol{h}_{nb}^{\sss (s)}),
\end{align*}
where $\boldsymbol{W}^{\sss (t_{\scalebox{0.5}{\textit{i}}})}$ and $\boldsymbol{b}^{\sss (t_{\scalebox{0.5}{\textit{i}}})}$ are the learnable parameters in the additional linear projection layer of the $i$-th teacher model.
The overall embedding loss is the weighted summation of all these embedding losses, i.e., $\mathcal{L}^{\sss \rm{emb}}_{\sss \rm{DP}}=\sum_{i=1}^{\sss M} \alpha^{\sss (t_{\scalebox{0.5}{\textit{i}}})}\mathcal{L}^{\sss {\rm{emb}_{\scalebox{0.5}{\textit{i}}}}}_{\sss \rm{DP}}$.
The loss function for the student model in Stage I is the summation of the distillation loss, the overall embedding loss, and its InfoNCE loss in our domain-specific post-training task, which is formulated as follows:
\begin{equation*}
  \mathcal{L}_{\sss 1}=\mathcal{L}^{\sss \rm{distill}}_{\sss \rm{DP}}+\mathcal{L}^{\sss \rm{emb}}_{\sss \rm{DP}}+\mathcal{L}_{\sss \rm{DP}}^{\sss (s)}.
\end{equation*}

Next, in Stage II, we first finetune these $M$ post-trained teacher news encoders with the news recommendation task (Step 3).
Then they are used to transfer rich task-specific knowledge to the student during its finetuning (Step 4).
Similar to Stage I, we assign a weight $\beta^{\sss (t_{\scalebox{0.5}{\textit{i}}})}$ to each finetuned teacher model based on its cross-entropy loss given an input sample of the news recommendation task and apply the following distillation loss to adjust the output of the student model during its finetuning:
\begin{gather*}
  \beta^{\sss (t_{\scalebox{0.5}{\textit{i}}})} = \frac{\exp(-\operatorname{CE}(\hat{\boldsymbol{y}}_{\sss \rm{FT}}^{\sss (t_{\scalebox{0.5}{\textit{i}}})}, y_{\sss \rm{FT}}))}{\sum_{j=1}^{\sss M}{\exp(-\operatorname{CE}(\hat{\boldsymbol{y}}_{\sss \rm{FT}}^{\sss (t_j)}, y_{\sss \rm{FT}}))}},\\
  \mathcal{L}^{\sss \rm{distill}}_{\sss \rm{FT}} = T_{\sss \rm{FT}}^{\sss 2}\cdot\operatorname{CE}(\sum_{i=1}^M{\beta^{\sss (t_{\scalebox{0.5}{\textit{i}}})}\hat{\boldsymbol{y}}_{\sss \rm{FT}}^{\sss (t_{\scalebox{0.5}{\textit{i}}})}}/T_{\sss \rm{FT}}, \hat{\boldsymbol{y}}_{\sss \rm{FT}}^{\sss (s)}/T_{\sss \rm{FT}}),
\end{gather*}
where $\hat{\boldsymbol{y}}_{\sss \rm{FT}}$ denotes the predicted score of the model on the news recommendation task and $T_{\sss \rm{FT}}$ is another temperature hyper-parameter.
We also use an additional embedding loss to align both the news representation and the user representation of the student model and the teacher models, which is formulated as follows:
\begin{align*}
  \begin{aligned}
    \mathcal{L}^{\sss \rm{emb}}_{\sss \rm{FT}} = \sum_{i=1}^{M} \beta^{\sss (t_{\scalebox{0.5}{\textit{i}}})}[ & \operatorname{MSE}(\boldsymbol{W}^{\sss (t_{\scalebox{0.5}{\textit{i}}})}_{n}\boldsymbol{n}^{\sss (t_{\scalebox{0.5}{\textit{i}}})}+\boldsymbol{b}^{\sss (t_{\scalebox{0.5}{\textit{i}}})}_{n}, \boldsymbol{n}^{\sss (s)})+  \\[-3mm]
                                                                                                               & \operatorname{MSE}(\boldsymbol{W}^{\sss (t_{\scalebox{0.5}{\textit{i}}})}_{u}\boldsymbol{u}^{\sss (t_{\scalebox{0.5}{\textit{i}}})}+\boldsymbol{b}^{\sss (t_{\scalebox{0.5}{\textit{i}}})}_{u}, \boldsymbol{u}^{\sss (s)})],
  \end{aligned}
\end{align*}
where $\boldsymbol{W}^{\sss (t_{\scalebox{0.5}{\textit{i}}})}_{n}$, $\boldsymbol{b}^{\sss (t_{\scalebox{0.5}{\textit{i}}})}_{n}$ and $\boldsymbol{W}^{\sss (t_{\scalebox{0.5}{\textit{i}}})}_{u}$, $\boldsymbol{b}^{\sss (t_{\scalebox{0.5}{\textit{i}}})}_{u}$ are the learnable parameters used to project the news representations and the user presentations learned by the $i$-th teacher model into a unified space, respectively.
The student model is also tuned to minimize the cross-entropy loss between its predicted score $\hat{\boldsymbol{y}}_{\sss {\rm{FT}}}^{\sss (s)}$ and the ground-truth label $y_{\sss \rm{FT}}$ of the news recommendation task, i.e., $\mathcal{L}_{\sss \rm{FT}}^{\sss (s)}=\operatorname{CE}(\hat{\boldsymbol{y}}_{\sss \rm{FT}}^{\sss (s)}, y_{\sss \rm{FT}})$.
The overall loss function for the student model in Stage II is the summation of the distillation loss, the embedding loss, and its finetuning loss, which is formulated as follows:
\begin{equation*}
  \mathcal{L}_{\sss 2}=\mathcal{L}^{\sss \rm{distill}}_{\sss \rm{FT}}+\mathcal{L}^{\sss \rm{emb}}_{\sss \rm{FT}}+\mathcal{L}_{\sss \rm{FT}}^{\sss (s)}.
\end{equation*}
\section{Experiments}

\subsection{Datasets and Experimental Settings}
\begin{table}[!t]
  \centering
  \setlength{\belowcaptionskip}{-5pt}
  \resizebox{\linewidth}{!}{
    \begin{tabular}{lrlr}
      \Xhline{1.5pt}
      \multicolumn{4}{c}{\textbf{MIND}}                               \\ \hline
      \# News           & 161,013    & \# Users          & 1,000,000  \\
      \# Impressions    & 15,777,377 & \# Clicks         & 24,155,470 \\
      Avg. title length & 11.52      &                   &            \\ \hline
      \multicolumn{4}{c}{\textbf{Feeds}}                              \\ \hline
      \# News           & 377,296    & \# Users          & 10,000     \\
      \# Impressions    & 320,925    & \# Clicks         & 437,072    \\
      Avg. title length & 11.93      &                   &            \\ \hline
      \multicolumn{4}{c}{\textbf{News}}                               \\ \hline
      \# News           & 1,975,767  & Avg. title length & 11.84      \\
      Avg. body length  & 511.43     &                   &            \\
      \Xhline{1.5pt}
    \end{tabular}
  }
  \caption{Detailed statistics of \textit{MIND}, \textit{Feeds} and \textit{News}.}\label{dataset}
\end{table}
We conduct experiments with three real-world datasets, i.e., \textit{MIND}, \textit{Feeds}, and \textit{News}.
\textit{MIND} is a public dataset for news recommendation~\cite{wu2020mind}, which contains the news click logs of 1,000,000 users on the Microsoft News website in six weeks.
We use its public training set, validation set, and test set for experiments\footnote{We randomly choose 1/2 samples from the original training set as our training data due to the limit of training speed.}.
\textit{Feeds} is also a news recommendation dataset collected on the Microsoft News App from 2020-08-01 to 2020-09-01.
We use the impressions in the last week for testing and randomly sampled 20\% impressions from the training set for validation.
\textit{News} contains news articles collected on the Microsoft News website from 2020-09-01 to 2020-10-01, which is used for our domain-specific post-training task.
Detailed statistics of these datasets are summarized in Table~\ref{dataset}.

\begin{table*}[t]
  \setlength{\belowcaptionskip}{-5pt}
  \resizebox{\linewidth}{!}{
    \begin{tabular}{l|lll|lll|c}
      \Xhline{1.5pt}
      \multicolumn{1}{c|}{\multirow{2}{*}{\textbf{Model}}} & \multicolumn{3}{c|}{\textbf{MIND}} & \multicolumn{3}{c|}{\textbf{Feeds}} & \multicolumn{1}{c}{\multirow{2}{*}{\textbf{\begin{tabular}[c]{@{}c@{}}Model \\ Size\end{tabular}}}}                                                                                                           \\ \cline{2-7}
      \multicolumn{1}{c|}{}                                & \multicolumn{1}{c}{AUC}            & \multicolumn{1}{c}{MRR}             & \multicolumn{1}{c|}{nDCG@10}                                                                        & \multicolumn{1}{c}{AUC} & \multicolumn{1}{c}{MRR} & \multicolumn{1}{c|}{nDCG@10} & \multicolumn{1}{c}{} \\ \hline
      PLM-NR$_{12}$ (FT)                                   & 69.72±0.15                         & 34.74±0.10                          & 43.71±0.07                                                                                          & 67.93±0.13              & 34.42±0.07              & 45.09±0.07                   & 109.89M              \\
      PLM-NR$_{12}$ (DAPT)                                 & 69.97±0.08                         & 35.07±0.15                          & 43.98±0.10                                                                                          & 68.24±0.09              & 34.63±0.10              & 45.30±0.09                   & 109.89M              \\
      PLM-NR$_{12}$ (TAPT)                                 & 69.82±0.14                         & 34.90±0.11                          & 43.83±0.07                                                                                          & 68.11±0.11              & 34.49±0.12              & 45.11±0.08                   & 109.89M              \\
      PLM-NR$_{12}$ (DP)                                   & \textbf{71.02±0.07}                & \textbf{36.05±0.09}                 & \textbf{45.03±0.12}                                                                                 & \textbf{69.37±0.10}     & \textbf{35.74±0.11}     & \textbf{46.45±0.11}          & 109.89M              \\ \hline
      PLM-NR$_4$ (FT)                                      & 69.49±0.14                         & 34.40±0.10                          & 43.40±0.09                                                                                          & 67.46±0.12              & 33.71±0.11              & 44.36±0.09                   & 53.18M               \\
      PLM-NR$_2$ (FT)                                      & 68.99±0.08                         & 33.59±0.14                          & 42.61±0.11                                                                                          & 67.05±0.14              & 33.33±0.09              & 43.90±0.12                   & 39.01M               \\
      PLM-NR$_1$ (FT)                                      & 68.12±0.12                         & 33.20±0.07                          & 42.07±0.10                                                                                          & 66.26±0.10              & 32.55±0.12              & 42.99±0.09                   & 31.92M               \\ \hline
      TinyBERT$_4$                                         & 70.55±0.10                         & 35.60±0.12                          & 44.47±0.08                                                                                          & 68.40±0.08              & 34.64±0.10              & 45.21±0.11                   & 53.18M               \\
      TinyBERT$_2$                                         & 70.24±0.13                         & 34.93±0.07                          & 43.98±0.10                                                                                          & 68.01±0.07              & 34.37±0.09              & 44.90±0.10                   & 39.01M               \\
      TinyBERT$_1$                                         & 69.19±0.09                         & 34.35±0.10                          & 43.12±0.07                                                                                          & 67.16±0.11              & 33.42±0.07              & 43.95±0.07                   & 31.92M               \\ \hline
      NewsBERT$_4$                                         & 70.62±0.15                         & 35.72±0.11                          & 44.65±0.08                                                                                          & 68.69±0.10              & 34.90±0.08              & 45.64±0.11                   & 53.18M               \\
      NewsBERT$_2$                                         & 70.41±0.09                         & 35.46±0.07                          & 44.35±0.10                                                                                          & 68.24±0.09              & 34.64±0.11              & 45.23±0.10                   & 39.01M               \\
      NewsBERT$_1$                                         & 69.45±0.11                         & 34.75±0.09                          & 43.54±0.12                                                                                          & 67.37±0.05              & 33.55±0.10              & 44.12±0.08                   & 31.92M               \\ \hline
      Tiny-NewsRec$_4$                                     & \textbf{71.19±0.08}                & \textbf{36.21±0.05}                 & \textbf{45.20±0.09}                                                                                 & \textbf{69.58±0.06}     & \textbf{35.90±0.11}     & \textbf{46.57±0.07}          & 53.18M               \\
      Tiny-NewsRec$_2$                                     & 70.95±0.04                         & 36.05±0.08                          & 44.93±0.10                                                                                          & 69.25±0.07              & 35.45±0.09              & 46.25±0.10                   & 39.01M               \\
      Tiny-NewsRec$_1$                                     & 70.04±0.06                         & 35.16±0.10                          & 44.10±0.08                                                                                          & 68.31±0.03              & 34.65±0.08              & 45.32±0.08                   & 31.92M               \\ \Xhline{1.5pt}
    \end{tabular}
  }
  \caption{Performance comparisons of different models. The results of the best-performed teacher model and student model are highlighted. The subscript number denotes the number of layers in the model. The model size is measured by the number of parameters.}\label{result}
\end{table*}

In our experiments, following PLM-NR~\cite{wu2021empower}, we apply the pre-trained UniLMv2 \cite{bao2020unilmv2} to initialize the PLM in the news encoder due to its superior text modeling capability.
The dimensions of the news representation and the user representation are both 256.
The temperature hyper-parameters $T_{\sss \rm{DP}}$ and $T_{\sss \rm{FT}}$ are both set to 1.
A copy of the teacher model is saved every $K=500$ steps during post-training and the number of teacher models $M$ is set to 4.
We use the Adam optimizer \cite{kingma2014adam} for training.
The detailed experimental settings are listed in the Appendix.
All the hyper-parameters are tuned on the validation set.
Following \citet{wu2020mind}, we use AUC, MRR, and nDCG@10 to measure the performance of news recommendation models.
We independently repeat each experiment 5 times and report the average results with standard deviations.

\subsection{Performance Comparison}
In this section, we compare the performance of the 12-layer teacher model PLM-NR$_{12}$ (DP) which is domain-specifically post-trained before finetuning, and the student models trained with our Tiny-NewsRec with the following baseline methods:
\begin{itemize}[leftmargin=*,itemsep=2pt,parsep=0pt,topsep=4pt]
  \item \textbf{PLM-NR (FT)}~\cite{wu2021empower}, the state-of-the-art PLM-based news recommendation method which applies the PLM to the news encoder and directly fine-tunes it with the news recommendation task. We compare the performance of its 12-layer version and its variant using the first 1, 2, or 4 layers of the PLM.
  \item \textbf{PLM-NR (DAPT)}, a variant of PLM-NR which first adapts the PLM to the news domain via domain-adaptive pre-training~\cite{gururangan2020adapt}. It continues to pre-train the PLM on a corpus of unlabeled news domain texts and then finetunes it with the news recommendation task.
  \item \textbf{PLM-NR (TAPT)}, another variant of PLM-NR which first adapts the PLM to the downstream task with task-adaptive pre-training~\cite{gururangan2020adapt}. It continues to pre-train the PLM on the unlabeled news set provided along with the downstream training data and then finetunes it with the news recommendation task.
  \item \textbf{TinyBERT} \cite{jiao2020tinybert}, a state-of-the-art two-stage knowledge distillation method for compressing the PLM which conducts knowledge distillation in both the pre-training stage and the finetuning stage. For a fair comparison, we use the PLM-NR$_{12}$ (DP) as the teacher model.
  \item \textbf{NewsBERT} \cite{wu2021newsbert}, a PLM knowledge distillation method specialized for intelligent news applications which jointly trains the student model and the teacher model during finetuning. For a fair comparison, we use the 12-layer domain-specifically post-trained news encoder to initialize the teacher model.
\end{itemize}

Table~\ref{result} shows the performance of all these methods on the \textit{MIND} and \textit{Feeds} datasets.
From the results, we have the following observations.
First, both PLM-NR$_{12}$ (DAPT) and PLM-NR$_{12}$ (TAPT) outperform PLM-NR$_{12}$ (FT).
It validates that continued pre-training on the corpus related to the downstream task can mitigate the domain shift problem to some extent.
Second, our PLM-NR$_{12}$ (DP) achieves the best performance among all 12-layer models.
This is because our proposed self-supervised domain-specific post-training task can help the PLM better capture the semantic information in news texts and generate more discriminative news representations, which is beneficial to the news understanding and user interest matching in the news recommendation task.
Third, compared with state-of-the-art knowledge distillation methods (i.e., NewsBERT and TinyBERT), our Tiny-NewsRec achieves the best performance in all 1-layer, 2-layer, and 4-layer student models, and our further t-test results show the improvements are significant at $p<0.01$ (by comparing the models with the same number of layers).
This is because the student model can better adapt to the news domain with supervision from the domain-specifically post-trained teacher models in Stage I, and task-specific knowledge is also transferred to it during the knowledge distillation in Stage II.
Finally, our Tiny-NewsRec even achieves comparable performance with the teacher model PLM-NR$_{12}$ (DP) while having much fewer parameters and lower computational overhead.
This is because these multiple teacher models originated from different time steps of the post-training procedure may complement each other and provide more comprehensive knowledge to the student model in both stages.

\subsection{Further Comparison}
\begin{table}[t]
  \setlength{\belowcaptionskip}{-5pt}
  \resizebox{\linewidth}{!}{
    \begin{tabular}{l|ccc}
      \Xhline{1.5pt}
      \multicolumn{1}{c|}{\textbf{Model}} & \multicolumn{1}{c}{\textbf{AUC}} & \multicolumn{1}{c}{\textbf{MRR}} & \multicolumn{1}{c}{\textbf{nDCG@10}} \\ \hline
      Ensemble-Teacher$_{12}$             & 69.43                            & 35.81                            & 46.53                                \\
      TinyBERT-MT$_4$                     & 68.87                            & 35.13                            & 45.81                                \\
      NewsBERT-MT$_4$                     & 68.82                            & 35.07                            & 45.80                                \\
      MT-BERT$_4$                         & 68.51                            & 34.74                            & 45.45                                \\
      Tiny-NewsRec$_4$                    & \textbf{69.58}                   & \textbf{35.90}                   & \textbf{46.57}                       \\  \Xhline{1.5pt}
    \end{tabular}
  }
  \caption{Performance comparisons of the ensemble teacher models and the student models distilled with various multi-teacher knowledge distillation methods.}\label{further}
\end{table}

To better understand where the performance improvement of our approach comes from, we further compare our Tiny-NewsRec with the following methods which use multiple teacher models:
\begin{itemize}[leftmargin=*,itemsep=2pt,parsep=0pt,topsep=4pt]
  \item \textbf{Ensemble-Teacher}, which is the ensemble of the multiple 12-layer teacher models used by Tiny-NewsRec. The average predicted score of these teacher models is used for evaluation.
  \item \textbf{TinyBERT-MT} and \textbf{NewsBERT-MT}, the modified version of TinyBERT~\cite{jiao2020tinybert} and NewsBERT~\cite{wu2021newsbert}, which utilize the multiple teacher models used by Tiny-NewsRec. Each teacher model is adaptively weighted according to its performance on the input training sample, which is the same as the one used in our two-stage knowledge distillation method.
  \item \textbf{MT-BERT}~\cite{wu2021teacher}, which jointly finetunes the student model and multiple teacher models with different PLMs on the downstream news recommendation task.
\end{itemize}

Table~\ref{further} shows the performance of the ensemble teacher models and the 4-layer student models on the \textit{Feeds} dataset.
Comparing with the results of PLM-NR$_{12}$ (DP) in Table~\ref{result}, we first find that the simple ensemble of multiple teacher models cannot bring much performance gain.
The reason is that these teachers are treated equally during testing.
However, in our Tiny-NewsRec, for each training sample, we assign an adaptive weight to each teacher model based on their performance.
The student can always learn more from the best teacher model on each sample and receive more comprehensive knowledge.
Second, even with the same teacher models, Tiny-NewsRec still outperforms TinyBERT-MT and NewsBERT-MT.
This is because we are the first to use multiple teacher models to transfer domain-specific knowledge to the student before the task-specific finetuning, which can help the student model better adapt to the news domain.
Besides, we find that MT-BERT achieves the worst performance among all the compared methods.
It verifies that the multiple teacher models originating from different time steps of our post-training procedure can provide more comprehensive knowledge than these jointly finetuned teacher models with different PLMs used in MT-BERT.

\subsection{Effectiveness of Multiple Teacher Models}\label{ablation_teacher}
\begin{figure}[t]
  \centering
  \setlength{\belowcaptionskip}{-5pt}
  \includegraphics[width=\linewidth]{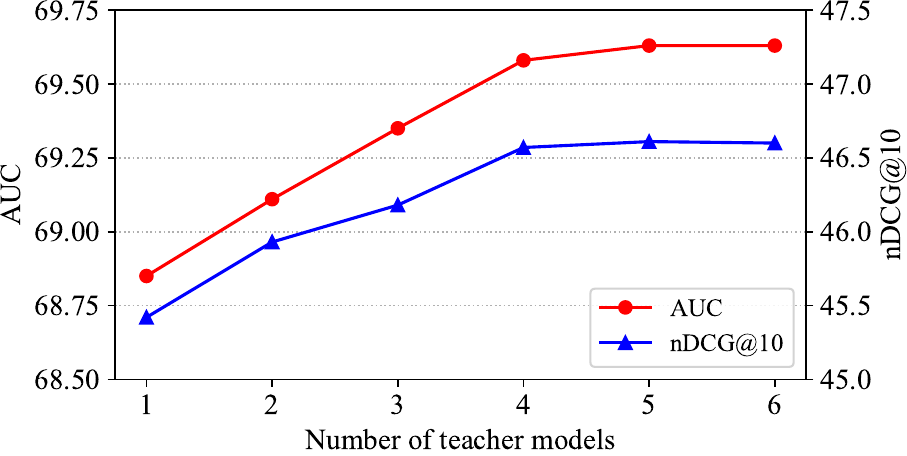}
  \caption{Impact of the number of teacher models $M$.}
  \label{fig.num_teacher}
\end{figure}
In this subsection, we conduct experiments to explore the impact of the number of teacher models in our Tiny-NewsRec.
We vary the number of teacher models $M$ from 1 to 6 and compare the performance of the 4-layer student model on the \textit{Feeds} dataset\footnote{The results on the \textit{MIND} dataset show similar trends and are placed in the Appendix.\label{appendix}}.
The results are shown in Fig.~\ref{fig.num_teacher}.
From the results, we find that the performance of the student model first greatly improves with the number of teacher models.
This is because these teacher models at different time steps of the post-training procedure usually can complement each other.
With more teacher models, the student model can receive more comprehensive knowledge and obtain better generalization ability.
However, increasing the number of teacher models can only bring marginal improvement when $M$ is larger than 4, which may reach the upper bound of the performance gain brought by the ensemble of multiple teachers.
Thus we set $M$ to 4 in our Tiny-NewsRec as a balance between the model performance and the additional training cost of these teacher models.

\subsection{Effectiveness of Two-stage Knowledge Distillation}
\begin{figure}[t]
  \centering
  \setlength{\belowcaptionskip}{-5pt}
  \includegraphics[width=\linewidth]{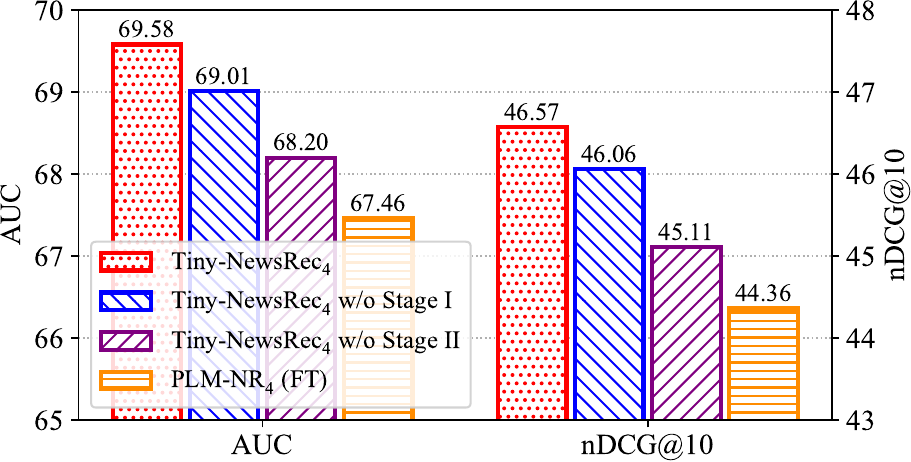}
  \caption{Effectiveness of each stage in our framework.}
  \label{fig.ablation}
\end{figure}
In this subsection, we further conduct several experiments to verify the effectiveness of each stage in our two-stage knowledge distillation method.
We compare the performance of the 4-layer student model distilled with our Tiny-NewsRec and its variant with one stage removed on the \textit{Feeds} dataset\textsuperscript{\ref{appendix}}.
The results are shown in Fig.~\ref{fig.ablation}.
From the results, we first find that the knowledge distillation in Stage II plays a critical role in our approach as the performance of the student model declines significantly when it is removed.
This is because the guidance from the teacher models in the second stage such as learned news and user representations can provide much more useful information than the one-hot ground-truth label, which encourages the student model to behave similarly to the teacher models in the news recommendation task.
The complement between multiple teacher models also enables the student model to achieve better generalization ability.
Second, the performance of the student model also declines after we remove Stage I.
This is because our self-supervised domain-specific post-training task can make the PLM better adapt to the news domain and generate more discriminative news representations.
The multiple teacher models can also transfer useful domain-specific knowledge to the student model during its domain adaptation.

\subsection{Effectiveness of Each Loss Function}
\begin{figure}[t]
  \centering
  \setlength{\belowcaptionskip}{-5pt}
  \includegraphics[width=\linewidth]{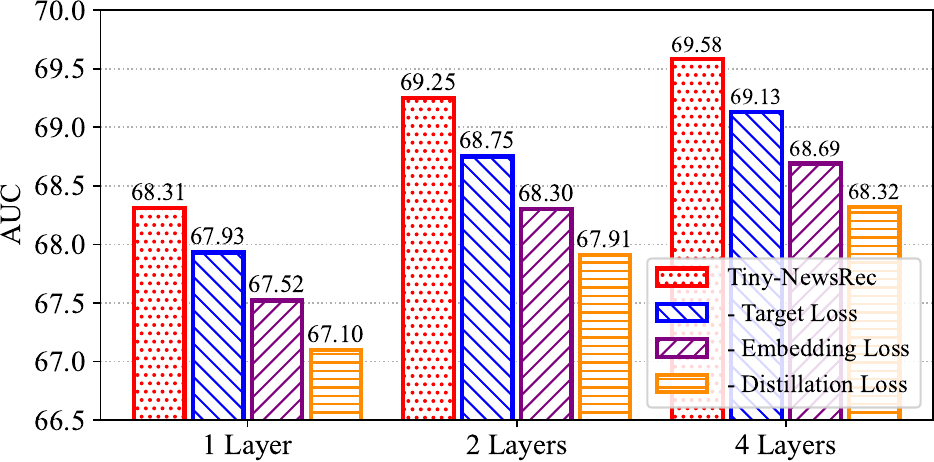}
  \caption{Effectiveness of each loss function.}
  \label{fig.loss}
\end{figure}
In this subsection, we further explore the impact of each part of the overall loss function in our two-stage knowledge distillation method, i.e., the distillation loss ($\mathcal{L}^{\sss \rm{distill}}_{\sss \rm{DP}}$ and $\mathcal{L}^{\sss \rm{distill}}_{\sss \rm{FT}}$), the embedding loss ($\mathcal{L}^{\sss \rm{emb}}_{\sss \rm{DP}}$ and $\mathcal{L}^{\sss \rm{emb}}_{\sss \rm{FT}}$), and the target loss ($\mathcal{L}_{\sss \rm{DP}}^{\sss (s)}$ and $\mathcal{L}_{\sss \rm{FT}}^{\sss (s)}$).
We compare the performance of the student models distilled with our Tiny-NewsRec approach and its variant with one part of the overall loss function removed.
The results on the \textit{Feeds} dataset are shown in Fig.~\ref{fig.loss}.
From the results, we have several findings.
First, the distillation loss is the most essential part of the overall loss function as the performance of the student model drops significantly after it is removed.
This is because the distillation loss can force the student model to make similar predictions as the teacher model, which directly decides the performance of the student model on the news recommendation task.
In addition, the embedding loss is also important in our approach.
It may be because the embedding loss aligns the news representations and the user representations learned by the student model and the teacher models, which can help the student model better imitate the teacher models.
Besides, the target loss is also useful for the training of the student model.
This may be because these finetuned teacher models will still make some mistakes in certain training samples.
The supervision from the ground-truth label is still necessary for the student model.

\subsection{Efficiency Evaluation}
\begin{figure}[t]
  \centering
  \setlength{\belowcaptionskip}{-5pt}
  \includegraphics[width=\linewidth]{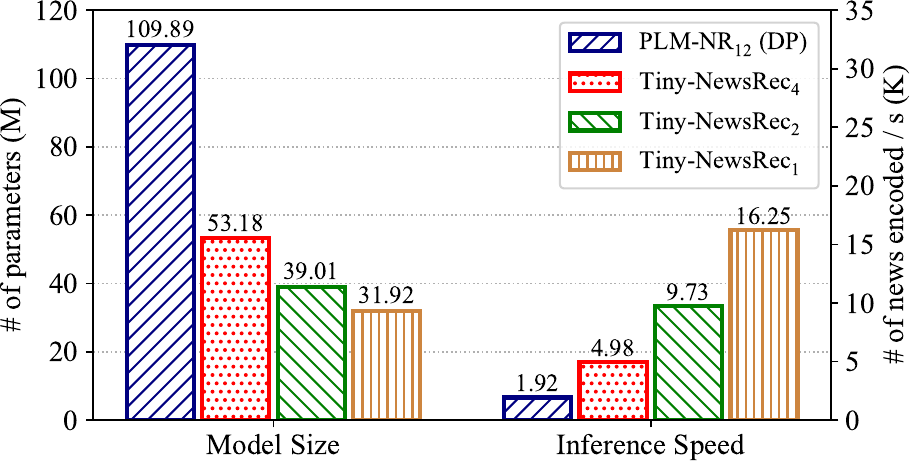}
  \caption{Model size and inference speed of the teacher model and student models.}
  \label{fig.speed}
\end{figure}
In this subsection, we conduct experiments to evaluate the efficiency of the student models distilled with our Tiny-NewsRec.
As in news recommendation, encoding news with the PLM-based news encoder is the main computational overhead, we measure the inference speed of the model in terms of the number of news that can be encoded per second with a single GPU.
We also measure the model size by the number of parameters.
The evaluation results of the 1-layer, 2-layer, and 4-layer student models and the 12-layer teacher model are shown in Fig.~\ref{fig.speed}.
The results show that our Tiny-NewsRec can reduce the model size by 50\%-70\% and increase the inference speed by 2-8 times while achieving better performance.
These results verify that our approach can improve the effectiveness and efficiency of the PLM-based news recommendation model at the same time.
It is noted that the student model distilled with other knowledge distillation methods (e.g., TinyBERT and NewsBERT) can achieve the same inference speed as Tiny-NewsRec since the structure of the final student model is the same.
However, our Tiny-NewsRec can get much better performance as shown in Table~\ref{result} and~\ref{further}.
\section{Conclusion}

In this paper, we propose a Tiny-NewsRec method to improve the effectiveness and efficiency of PLM-based news recommendation with domain-specific post-training and a two-stage knowledge distillation method. 
Specifically, before the task-specific finetuning, we propose to conduct domain-specific post-training towards the PLM-based news encoder with a self-supervised matching task between news titles and news bodies to make the general PLM better capture and match the high-level semantic information in news texts.
In our two-stage knowledge distillation method, the student model is first adapted to the news domain and then finetuned on the news recommendation task with the domain-specific and task-specific knowledge transferred from multiple teacher models in each stage.
We conduct extensive experiments on two real-world datasets and the results demonstrate that our approach can effectively improve the performance of the PLM-based news recommendation model with considerably smaller models.
\section{Ethics Statement}
In this paper, we conduct experiments with three real-world datasets, i.e., \textit{MIND}, \textit{Feeds}, and \textit{News}.
\textit{MIND} is a public English news recommendation dataset released in~\cite{wu2020mind}.
In this dataset, each user was delinked from the production system when securely hashed into an anonymized ID using one-time salt mapping to protect user privacy.
We have agreed to Microsoft Research License Terms\footnote{\href{https://msnews.github.io/}{https://msnews.github.io/}} before downloading the dataset.
\textit{Feeds} is another news recommendation dataset collected on the Microsoft News App.
It followed the same processing procedure as \textit{MIND}, using the one-time salt mapping to securely hash each user into an anonymized ID.
\textit{News} is a news article dataset collected on the Microsoft News website which only contains public news articles and no user-related information is involved.
Thus, all the datasets used in our paper will not reveal any user privacy information.
\section{Limitations}
In our Tiny-NewsRec, we utilize multiple teacher models to transfer comprehensive knowledge to the student model in our two-stage knowledge distillation method.
These teacher models originate from different time steps of the post-training procedure and later they are finetuned with the news recommendation task separately.
Our ablation study verifies the effectiveness of multiple teacher models.
However, training a teacher model requires lots of time and computing resources as it contains a large PLM.
Compared with existing single-teacher knowledge distillation methods, our approach will enlarge the training cost by $M$ times in order to obtain $M$ high-quality teacher models.
We will try to reduce the training cost of our approach while keeping its performance in our future work.
\section{Acknowledgements}

This research was partially supported by grants from the National Key Research and Development Program of China (No. 2021YFF0901003), and the National Natural Science Foundation of China (No. 61922073 and U20A20229).

\bibliography{anthology,custom}
\bibliographystyle{acl_natbib}

\appendix
\clearpage
\section{Appendix}
\subsection{Experimental Settings}
In our domain-specific post-training, we use the first 24 tokens of the news title and the first 512 tokens of the news body for news title and news body modeling. We use the pre-trained UniLMv2 model as the PLM and only finetune its last three Transformer layers.
During finetuning with the news recommendation task, we use the first 30 tokens of the news title for news modeling. We also use the UniLMv2 model as the PLM and only finetune its last two Transformer layers as we find that finetuning all the parameters does not bring significant gain in model performance but drastically slows down the training speed.
The complete hyper-parameter settings are listed in Table \ref{hyper-param}.

\subsection{Additional Results on \textit{MIND}}
\begin{figure}[!t]
    \centering
    \includegraphics[width=\linewidth]{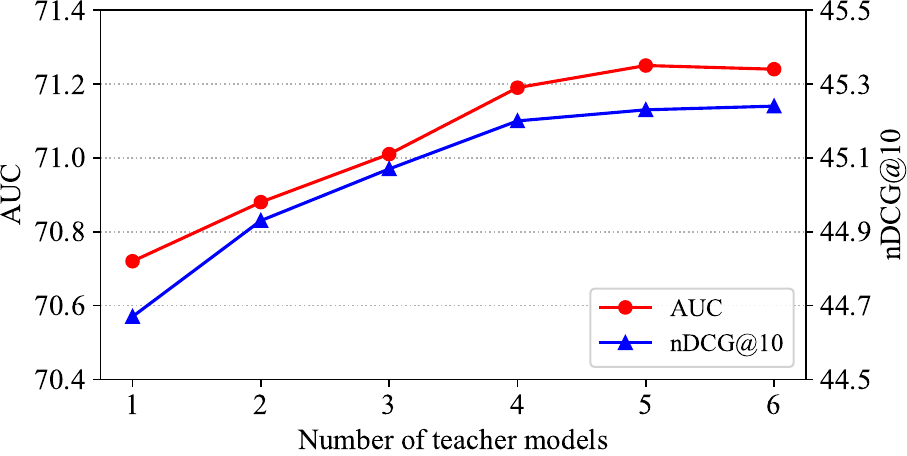}
    \caption{Impact of the number of teacher models $M$.}
    \label{fig.mind_teacher}
\end{figure}

\begin{figure}[!t]
    \centering
    \includegraphics[width=\linewidth]{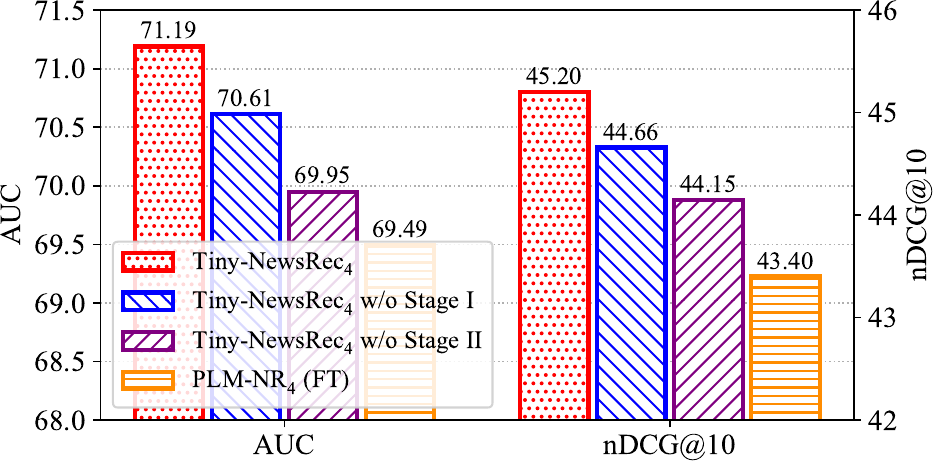}
    \caption{Effectiveness of each stage in our framework.}
    \label{fig.ablation_mind}
\end{figure}

We also report the additional results on the \textit{MIND} dataset, which are shown in Fig.~\ref{fig.mind_teacher} and Fig.~\ref{fig.ablation_mind}.
We observe phenomena similar to the results on the \textit{Feeds} dataset.

\subsection{Experimental Environment}
We conduct experiments on a Linux server with Ubuntu 18.04.1.
The server has 4 Tesla V100-SXM2-32GB GPUs with CUDA 11.0.
The CPU is Intel(R) Xeon(R) Platinum 8168 CPU @ 2.70GHz and the total memory is 661GB.
We use Python 3.6.9 and PyTorch 1.6.0.
In our domain-specific post-training and the post-training stage knowledge distillation experiments, the model is trained on a single GPU. All the other models are parallelly trained on 4 GPUs with the Horovod framework.

\subsection{Running Time}
On the \textit{News} dataset, the domain-specific post-training of the 12-layer teacher model and the post-training stage knowledge distillation of the 4-layer, 2-layer, and 1-layer student models takes around 12 hours, 10 hours, 8 hours, and 6 hours respectively with a single GPU. On the \textit{MIND} dataset, the fine-tuning of the 12-layer teacher model and the finetuning stage knowledge distillation of the 4-layer, 2-layer, and 1-layer student models takes around 12 hours, 10 hours, 8 hours, and 6 hours respectively with 4 GPUs, while on the \textit{Feeds} dataset, it takes 3 hours, 2.5 hours, 2 hours, and 1.5 hours respectively.

\begin{table}[!t]
    \centering
    \resizebox{\linewidth}{!}{
        \begin{tabular}{cc}
            \Xhline{1.5pt}
            \multicolumn{2}{c}{\textbf{General Hyper-parameters}}                              \\ \hline
            \multicolumn{1}{c|}{Dimension of query vector in attention network} & 200          \\
            \multicolumn{1}{c|}{Adam betas}                                     & (0.9, 0.999) \\
            \multicolumn{1}{c|}{Adam eps}                                       & 1e-8         \\ \hline
            \multicolumn{2}{c}{\textbf{Domain-specific Post-training}}                         \\ \hline
            \multicolumn{1}{c|}{Negative sampling ratio $N$}                    & 9            \\
            \multicolumn{1}{c|}{Dimension of news title/body representation}    & 256          \\
            \multicolumn{1}{c|}{Batch size}                                     & 32           \\
            \multicolumn{1}{c|}{Learning rate}                                  & 1e-6         \\ \hline
            \multicolumn{2}{c}{\textbf{News Recommendation Finetuning}}                        \\ \hline
            \multicolumn{1}{c|}{Negative sampling ratio $S$}                    & 4            \\
            \multicolumn{1}{c|}{Max number of historical clicked news $L$}      & 50           \\
            \multicolumn{1}{c|}{Dimension of news/user representation}          & 256          \\
            \multicolumn{1}{c|}{Batch size}                                     & 32$\times$4  \\
            \multicolumn{1}{c|}{Learning rate}                                  & 5e-5         \\ \hline
            \multicolumn{2}{c}{\textbf{Two-stage Knowledge Distillation}}                      \\ \hline
            \multicolumn{1}{c|}{Temperature $T_{\sss \rm{DP}}$}                 & 1            \\
            \multicolumn{1}{c|}{Temperature $T_{\sss \rm{FT}}$}                 & 1            \\
            \multicolumn{1}{c|}{Number of teacher models $M$}                   & 4            \\
            \Xhline{1.5pt}
        \end{tabular}
    }
    \caption{Hyper-parameter settings}
    \label{hyper-param}
\end{table}


\end{document}